# Time asymmetry in ion diffusion under magnetic field

Ali Eftekhari

*Electrochemical Research Center, PO Box 19395-5139, Tehran, Iran*

**Abtract**

The general equation for the flux of an electrolyte in solution in the presence of an external magnetic field was derived mathematically in accordance with the Onsager formalism of irreversible thermodynamics. The time reversal symmetry was examined theoretically for the equation. Time reversal symmetry breaking was demonstrated for the case under investigation. Indeed, it was demonstrated theoretically that diffusion of an electrolyte is solution under an applied magnetic field has asymmetry in respect with time. The important of the study is to show time asymmetry in a physical process based on mathematical derivation of available equations.

* Corresponding author. Tel.: +98-21-204-2549; fax: +98-21-205-7621.

*E-mail address:* eftekhari@elchem.org.

# 1. Introduction

Time asymmetry is a conversial subject of research in modern physics. The concepts of time asymmetry in different physical systems have been extensively addressed in two books devoted to this subject [1,2]. The time asymmetry was a puzzling problem in the late nineteen century born from thought of irreversibility made by some scientists (mainly Boltzmann) [3-7]. This was a confusing problem in the light of the apparent time symmetry of the underlying laws of mechanics. The problem was that none of the mechanic versions including both classical and quantum mechanics could explain time asymmetry. In fact, all of dynamical laws are symmetric in respect with time (direction of time). This comes from the fact that current dynamical laws have been proposed based on the assumption of reversibility.

Time asymmetry can only be explained based on statistical mechanics and the second law of thermodynamics. As the occurrence of processes is accompanied with increase of entropy, if we change the direction of time to back, entropy should still be increased. Thus, processes are no longer symmetric in respect with time. Prigogine has used irreversible thermodynamics to explain time symmetry breaking [8]. Indeed, although, all versions of dynamics suggest time reversal symmetry or reversibility in respect with time, the statistical nature of the second law of thermodynamics is just able to explain time asymmetry. Unfortunately, this approach is not well established in the literature, as the time asymmetry is usually decreased by speculative statements based on the second law of thermodynamics. The published materials on the subject of time asymmetry are mainly restricted to descriptive book chapters published in the books devoted to with the subject of time [9-11]. Although, due to the importance of time asymmetry, it has been well documented in the literature in the field of philosophy of science [12-16]; however, less attention has been made to mathematical formulation of time asymmetry in physical processes, which is of interest in this area of research. Diffusion processes are well known to applied mathematicians, as they have widely used them as physical examples to propose mathematical models. In the present communication, we would like to investigate the existence of time asymmetry in diffusing ion motions under magnetic fields. As the mathematical relations were derived from fundamental concepts, the model proposed can simply show the time reversal symmetry breaking. Another interesting feature of the mathematical model is that on the contrary of previous works, the present one is not based on the concept of increasing entropy.

## 2. General relations for diffusion under magnetic field

The dissipation function has been well known for different cases [17]. In a general case, it can be expressed as:

$$\sigma = \sum_i J_j X_i \qquad (1)$$

where $J_i$ is flux and $X_i$ is its corresponding force. For a diffusing system, this force is referred to the gradient of chemical potential. Thus, we can express the dissipation function of a diffusing charged particle $\nu$ (anion or cation) with charge of $z_\pm$ as

$$\sigma_0 = \sum_i J_i \cdot grad(-\mu_i) \qquad (2)$$

In this case, the flux $J_i$ denotes the number of species $i$ passing through a unit area per unit time in a given direction. If we try to write the dissipation function for an applied magnetic field, we will have

$$\sigma_m = \sum_k J_k \cdot F_k \qquad (3)$$

where $F_i$ is the conjugate force of magnetic field, which can be defined as a Lorentz force:

$$F_i = z_i(V_i \times B) \qquad (4)$$

where $V_i$ denotes average drift velocity and $B$ the magnetic induction. Now, we can expressed the total dissipation function for a diffusing system under magnetic field as:

$$\sigma_t = \sigma_0 + \sigma_m \qquad (5)$$

and consequently

$$\sigma_t = -\sum_i J_i \cdot grad\ \mu_i + \sum_i J_i \cdot z_i(V_i \times B) \qquad (6)$$

For simplification, we restrict our investigation to a simple two-component system. Thus, we have

$$\sigma_t = \sum_{i=1}^{2} J_i \cdot [z_i(V_i \times B) - grad\ \mu_i] \qquad (7)$$

or

$$\sigma_t = J_1 \cdot [z_1(V_1 \times B) - grad\ \mu_1] + J_2 \cdot [z_2(V_2 \times B) - grad\ \mu_2] \qquad (8)$$

According to the Onsager reciprocity relations [18], we can express the fluxes of ions using Onsager's phenomenological coefficient:

$$J_1 = L_{11}[z_1(V_1 \times B) - grad\ \mu_1] + L_{12}[z_2(V_2 \times B) - grad\ \mu_2] \qquad (9)$$

$$J_2 = L_{21}[z_1(V_1 \times B) - grad\ \mu_1] + L_{22}[z_2(V_2 \times B) - grad\ \mu_2] \qquad (10)$$

As there is no electric current in the system, based on the conservation law, we have

$$z_1 J_1 + z_2 J_2 = 0 \qquad (12)$$

and consequently,

$$z_1\{L_{11}[z_1(V_1 \times B) - grad\ \mu_1] + L_{12}[z_2(V_2 \times B) - grad\ \mu_2]\} + \\ z_2\{L_{21}[z_1(V_1 \times B) - grad\ \mu_1] + L_{22}[z_2(V_2 \times B) - grad\ \mu_2]\} = 0 \qquad (13)$$

As the chemical potential of the electrolyte (salt) is equal to the sum of the separate chemical potentials of the ions:

$$grad\ \mu_s = v_1\ grad\ \mu_1 + v_2\ grad\ \mu_2 \qquad (14)$$

Thus, grad $\mu_1$ and grad $\mu_2$ can be expressed in terms of grad $\mu_s$:

$$grad\ \mu_1 = \frac{grad\ \mu_s(-z_1 L_{12} - z_2 L_{22}) + v_2(V_1 \times B)(z_1^2 L_{11} + z_1 z_2 L_{21}) + v_2(V_2 \times B)(z_1 z_2 L_{12} + z_2^2 L_{22})}{v_2(z_1 L_{11} + z_2 L_{21}) - v_1(z_1 L_{12} + z_2 L_{22})} \qquad (15)$$

$$grad\ \mu_2 = \frac{grad\ \mu_s(-z_1 L_{11} - z_2 L_{21}) + v_1(V_1 \times B)(z_2^2 L_{11} + z_1 z_2 L_{21}) + v_1(V_2 \times B)(z_1 z_2 L_{12} + z_2^2 L_{22})}{v_2(z_1 L_{11} + z_2 L_{21}) - v_1(z_1 L_{12} + z_2 L_{22})} \qquad (16)$$

Therefore, we can rewrite the fluxes in terms of grad $\mu_s$:

$$J_1 = L_{11}\left[\begin{array}{l} z_1(V_1 \times B) + \\ \dfrac{grad\ \mu_s(z_1 L_{12} + z_2 L_{22}) - v_2(V_1 \times B)(z_1^2 L_{11} + z_1 z_2 L_{21}) - v_2(V_2 \times B)(z_1 z_2 L_{12} + z_2^2 L_{22})}{v_2(z_1 L_{11} + z_2 L_{21}) - v_1(z_1 L_{12} + z_2 L_{22})} \end{array}\right]$$

$$+ L_{12}\left[\begin{array}{l} z_2(V_2 \times B) - \\ \dfrac{grad\ \mu_s(z_1 L_{11} + z_2 L_{21}) - v_1(V_1 \times B)(z_1^2 L_{11} + z_1 z_2 L_{21}) - v_2(V_2 \times B)(z_1 L_{12} + z_2^2 L_{22})}{v_2(z_1 L_{11} + z_2 L_{21}) - v_1(z_1 L_{12} + z_2 L_{22})} \end{array}\right] \qquad (17)$$

And

$$J_2 = L_{21}\left[\begin{array}{l} z_1(V_1 \times B) + \\ \dfrac{grad\ \mu_s(z_1 L_{12} + z_2 L_{22}) - v_2(V_1 \times B)(z_1^2 L_{11} + z_1 z_2 L_{21}) + v_2(V_2 \times B)(z_1 z_2 L_{12} + z_2^2 L_{22})}{v_2(z_1 L_{11} + z_2 L_{21}) - v_1(z_1 L_{12} + z_2 L_{22})} \end{array}\right]$$

$$+ L_{22}\left[\begin{array}{l} z_2(V_2 \times B) - \\ \dfrac{grad\ \mu_s(z_1 L_{11} + z_2 L_{21}) - v_1(V_1 \times B)(z_1^2 L_{11} + z_1 z_2 L_{21}) - v_2(V_2 \times B)(z_1 L_{12} + z_2^2 L_{22})}{v_2(z_1 L_{11} + z_2 L_{21}) - v_1(z_1 L_{12} + z_2 L_{22})} \end{array}\right] \qquad (18)$$

As the flux of the electrolyte can be expressed in terms of each components:

$$J_s = \frac{J_1}{v_1} = \frac{J_2}{v_2} \qquad (19)$$

we can write the flux of an electrolyte in the solution in the presence of magnetic field as:

$$J_s = \frac{L_{11}}{v_1} \times \left[ z_1(V_1 \times B) - \frac{v_2(V_1 \times B)(z_1^2 L_{11} + z_1 z_2 L_{12}) + v_2(V_2 \times B) z_1 z_2 L_{12} + z_2^2 L_{22}}{v_2(z_1 L_{11} + z_2 L_{21}) - v_1(z_1 L_{12} + z_2 L_{22})} \right]$$

$$+ \frac{L_{12}}{v_1} \left[ z_2(V_2 \times B) + \frac{v_1(V_1 \times B)(z_1^2 L_{11} + z_1 z_2 L_{21}) + v_1(V_2 \times B)(z_1 z_2 L_{12} + z_2^2 L_{22})}{v_2(z_1 L_{11} + z_2 L_{21}) - v_1(z_1 L_{12} + z_2 L_{22})} \right] \quad (20)$$

$$+ \operatorname{grad} \mu_s \left[ \begin{array}{c} \left(\dfrac{L_{11}}{v_1}\right) \dfrac{(z_1 L_{11} + z_2 L_{21})}{v_2(z_1 L_{11} + z_2 L_{21}) - v_1(z_1 L_{12} + z_2 L_{22})} \\ -\left(\dfrac{L_{12}}{v_1}\right) \dfrac{(z_1 L_{12} + z_2 L_{22})}{v_2(z_1 L_{11} + z_2 L_{21}) - v_1(z_1 L_{12} + z_2 L_{22})} \end{array} \right]$$

This can be considered as the general equation of the influence of an applied magnetic field on the flux of an electrolyte in the solution. It should be taken into account that this equation is valid for the assumption of the independency of the electrical conductance $K$ and the transference number of the solution from the applied magnetic field. For strong magnetic field, the validity of these assumptions is questionable and to be checked experimentally. However, we can use this equation for the most investigations using usual magnetic fields.

### 3. A typical electrolyte diffusion

As stated above, to simplify this equation, we restrict our investigation to a simple case. With the assumption of electro-neutrality of the solution, we have

$$v_1 z_1 + v_2 z_2 = 0 \quad (21)$$

For a simple case such as

$$KCl = K^+ + Cl^- \quad (22)$$

we have

$$z_1 = -z_2 = 1 \quad \text{and} \quad ; v_1 = v_2 = 1 \quad (23)$$
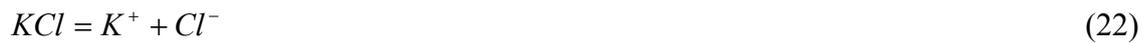

Thus, we can simplify the general equation Eq.(20) to the following one for the typical case:

$$J_s = \frac{L_{11}L_{22} - L_{12}^2}{(L_{11} - 2L_{12} + L_{22})}[(V_1 \times B) - (V_2 \times B) - grad\ \mu_s] \qquad (24)$$

It can be simplified by introducing a combined phenomenological coefficient $X$:

$$J_s = X[(V_1 \times B) - (V_2 \times B) - grad\ \mu_s] \qquad (25)$$

where $X = \dfrac{(L_{11}L_{22} - L_{12}^2)}{(L_{11} - 2L_{12} + L_{22})}$ \qquad (26)

We know that

$$grad\ \mu_s = \mu_{ss}\ grad\ c_s$$
$$\mu_{ss} = \partial \mu_s / \partial c_s \qquad (27)$$

Thus, the flux equation for the electrolyte in the solution for the typical example can be expressed as:

$$J_s = X[(V_1 \times B) - (V_2 \times B) - \mu_{ss}\ grad\ c_s] \qquad (28)$$

## 4. Time asymmetry aspect

Now we examine the existence of time reversal symmetry for the diffusion process of an electrolyte in the solution (the typical system introduced and similar systems) in the presence and absence of an applied magnetic field. To this aim, we investigate the time reversal symmetry for our simple two-component system, as expressed by Eq.(28). If the magnetic field vanishes, i.e. $\boldsymbol{B} \to 0$, in the absence of any external magnetic field, the equation Eq.(28) can be reduced to:

$$J_s^0 = -X\mu_{ss}\ grad\ c_s \qquad (29)$$

Similarly, according to the Fick's law, we have

$$J_s^0 = -D^0 \, grad \, c_s \tag{30}$$

Simply, it can be shown that Eq.(29) or Eq.(30) is symmetric in respect with time reversal. Under time reversal, when $t$ is changed to $-t$, the sign of flux will also be changed ($J \rightarrow -J$). On the other hand, we know that the diffusion coefficient is related to the mean frequency of actions [19,20] and under reversal time, the sign of the diffusion coefficient will also be changed ($D \rightarrow -D$). Moreover, if we investigate the value of the diffusion coefficient in accordance with the Stokes-Einstein equation:

$$D = kT / 6\pi\eta r \tag{31}$$

where $k$ is Boltzmann constant, $T$ the temperature, $\eta$ viscosity and $r$ the radii of the moving particle. It can be understood that no parameter in the equation is changed under time reversal. This means that the value of the diffusion coefficient when time direction is towards forward and backward is the same. Now, we divide the equation related to forward and backward time directions by using signs of "$\rightarrow$" for the forward time and "$\leftarrow$" for the backward time. All previously noted equation can be considered with the sign of "$\rightarrow$", as are based on usual forward arrow of time. Thus, we can rewrite Eq.(30) for the backward time as:

$$\overleftarrow{J_s^0} = -\overleftarrow{D^0} \, grad \, c_s \tag{32}$$

where $\overleftarrow{J_s^0} = -\overrightarrow{J_s^0}$ and $\overleftarrow{D^0} = -\overrightarrow{D^0}$. This shows that flux equation of an electrolyte in the solution (in the absence of an external magnetic field) has time reversal symmetry. Similar results can be obtained for Eq.(29) as the phenomenological coefficient (indicated as $X$) for a diffusing system is related to the diffusion coefficient.

Up to now, the results are in agreement of dynamical laws, which suggest time reversal symmetry. Now, we attempt to investigate the validity of time reversal symmetry for the flux of an electrolyte in the solution when an external magnetic field is applied. Let discuss the time reversal symmetry on the derived general equation of the flux of electrolytes in the presence of applied magnetic fields (Eq.(28)). As expected (similar to all dynamical relations, e.g. Eq.(29) or Eq.(30)), when $t \to -t$, the sign of flux will be changes $J \to -J$. In accordance with the change of the sign of diffusion coefficient ($D \to -D$), the phenomenological coefficient will also be changed its sign ($L \to -L$ or $X \to -X$) by changing the direction of time. Thus, to achieve time reversal symmetry for the flux of electrolytes in solution under an applied magnetic field, the sentence written in the parentheses (in Eq.(28)) should remain unchanged by changing the direction of time. Although, the concentration gradient is independent of time, the term related to the magnetic field seems to be dependent on the direction of time.

To investigate this problem, let investigate the basic concepts of time reversal [1] on the Lorentz equation (Eq.(4)). The concepts of this hypothesis has been discussed in [1]. Upon this action, the equation Eq.(4) will be changed to the following one:

$$\overrightarrow{F} = zV \times B \quad for\ t$$
$$\overleftarrow{F} = -zV \times B \quad for\ -t \tag{33}$$

Therefore, we should substitute $\overrightarrow{F}$ (the common F noted in Eq.(4) and other conventional equations) with $\overleftarrow{F}$ to express flux under backward direction of time, as it seems that $\overrightarrow{F} \neq \overleftarrow{F}$. Consequently, the general equation derived for the flux of an electrolyte in solution under an applied magnetic field changes under time reversal to:

$$-\overleftarrow{J_s} = -X\left[(-V_1 \times B) - (-V_2 \times B) - \mu_{ss}\ grad\ c_s\right] \tag{34}$$

By comparison of Eq.(28) and Eq.(34), it is obvious that

$$\overrightarrow{J_s} \neq -\overleftarrow{J_s} \tag{35}$$

which is indicative of time reversal asymmetry.

As seen, the equations mathematically derived show that the flux of an electrolyte in solution in the presence of an external magnetic field under forward direction of time is no longer equation to its flux under backward direction of time (Eq.(35)). Whereas, based on statistical mechanics or the second law of thermodynamics, we simply can speculate that as the forward flux ($\vec{J_s}$) is accompanied with increase of entropy, the backward flux should be accompanied with decrease of entropy to achieve time reversal symmetry. As based on the second law of thermodynamics processes cannot proceed spontaneously when entropy is decreases, thus, the entropy will also be increased even for the backward flux. Based on this speculative statement, we can claim time reversal symmetry breaking or time reversal asymmetry.

## 5. Basic electrodynamic

The results obtained mathematically can also be understood in Maxwell's equations. According to Faraday's induction law, which is applicable for the case of the present study:

$$\nabla \times \mathbf{E} = -\partial \mathbf{B} / \partial t \tag{36}$$

It is possible to write the induced electromotive force (emf) in terms of the vector potential **A** as the following integral form:

$$emf = \oint \mathbf{E} \cdot d\mathbf{s} - (d/dt) \iint \mathbf{B} \cdot d\mathbf{a} = -(d/dt) \iint (\nabla \times \mathbf{A}) \cdot d\mathbf{a} = -(d/dt) \oint \mathbf{A} \cdot d\mathbf{s} \tag{37}$$

suggesting

$$\mathbf{B}_{ind} = -d\mathbf{A}/dt \tag{38}$$

By keeping the integration region stationary, the induced electric field Eind containing total time derivative can be expressed as the partial time derivative. The incompleteness of the traditional formula for the total derivative of a vector field is:

$$dA/dt = \partial A/\partial t + (V \cdot \nabla)A \quad (39)$$

and the time range of change by a point moving with velocity **V** in a vector field **A** is expressed as [21]:

$$dA/dt = \partial A/\partial t + (V \cdot \nabla)A + (A \cdot \nabla)v \quad (40)$$

Although the vector identity:

$$(V \cdot \nabla)A + (A \cdot \nabla)V + A \times (\nabla \times V) - \nabla(V \cdot A) = -V \times (\nabla \times A) = -V \times B \quad (41)$$

For **V.A** = const. And $\nabla \times V = 0$, the 150 years old equation of Neumann is obtained:

$$E_{ind} = -\partial A/\partial t + V \times B \quad (42)$$

Now, time reversal symmetry can be inspected in Eq.(42). According to the hypothesis made in Eq.(33), the latter equation has not symmetry upon time reversal due to the existence of the term $V \times B$. Thus, we can conclude that $\vec{E}_{ind} \neq \overleftarrow{E}_{ind}$. This means that time asymmetry of the process of ion diffusion in an electrolyte solution upon time reversal is also confirmed by basic electrodynamic laws (Maxwell's equations).

## 6. Conclusion

As we are unable to perform any experiment to investigate time reversal symmetry (or asymmetry), this area of research is exclusively laid in theoretical physics. Nevertheless, such speculative statements are not the subject of extensive investigations for theoretical physicists (particularly mathematical physicists). Whereas, the present work showed that time asymmetry could be understood based on mathematical derivation of physical equations. The model proposed here can be used as a typical example to investigate time reversal symmetry of other physical equations. The validity of the results are obvious as a known hypothesis (described in [1]) was employed to the mathematical equations of magnetohydrodynamics. In fact, the importance of the present communication is to present a typical mathematical model showing time asymmetry and the approach employed here can be used for the investigation time asymmetry of various physical processes.